\documentstyle[aps,pra,multicol,epsf]{revtex}

\newcommand{\ket}{\rangle}
\newcommand{\nn}{\nonumber}
\newcommand{\uniE}{\hat{\bv{e}}}
\newcommand{\uniN}{\hat{\bv{n}}}
\newcommand{\uniM}{\hat{\bv{m}}}
\newcommand{\uniL}{\hat{\bv{l}}}

\newcommand{\nidanHajime}{\begin{multicols}{2}
  \global\columnwidth20.5pc\noindent}
\newcommand{\nidanOwari}{\end{multicols}\global\columnwidth42.5pc}
\global\columnwidth42.5pc
\begin{document}
\newcommand{\bv}[1]{\mbox{\boldmath$#1$}}
\draft
\title{
   Creation of persistent current and vortex in a 
   Bose-Einstein condensate of alkali-metal atoms
}
\author{
   Tomoya Isoshima\thanks{Electronic address: tomoya@mp.okayama-u.ac.jp},
   Mikio Nakahara\footnotemark[2], Tetsuo Ohmi\footnotemark[3], and
   Kazushige Machida
}
\address{
   Department of Physics, Okayama University, Okayama 700-8530, Japan
}
\address{
   \footnotemark[2]Department of Physics, Kinki University, Higashi-Osaka 577-8502, Japan
}
\address{
   \footnotemark[3]Department of Physics, Graduate School of Science,
   Kyoto University, Kyoto 606-8502, Japan
}
\date{\today}
\maketitle
\begin{abstract}
It is shown theoretically that a
persistent current can be continuously created in a 
Bose-Einstein condensate (BEC) of alkali atoms confined in 
a multiply connected region by making use of a spin-degree 
of freedom of the order parameter of a BEC.
We demonstrate that this persistent current is easily transformed into a vortex. 
Relaxation processes of these BEC after the 
confining field is turned off are also studied so that our 
analyses are compared with time of flight experiments.
The results are shown to clearly reflect the existence of a persistent current. 
\end{abstract}

\nidanHajime
\section{Introduction}

Since the discovery of Bose-Einstein condensation (BEC) in 
alkali atoms\ \cite{dalfovoReview}, numerous attempts are 
made to show that the system exhibits superfluidity.
One of these attempts is to create and observe quantized
vortices by confining the system in a rotating anisotropic trap.
Recently the vortex is created\ \cite{theVortex}
using the two-component BEC.

In the present paper, we propose a novel method to create a 
state with a persistent current or a vortex, where the hyperfine spin $F$
of alkali atoms is fully utilized.
This method has been briefly reported in\ \cite{LTabst},
and more detailed theoretical and numerical 
analyese are made in the present paper.
Although we restrict ourselves mainly to
the case $F=1$ to simplify our discussions, our method is also applicable
to cases with an arbitrary $F$.

BEC with $F=1$ may be expressed in terms of a three-component
order parameter, similarly to the spin or the
orbital part of superfluid $^3 $He.
In particular, a spin-polarized BEC has the same order
parameter as that of $^3$He-A\ \cite{woelfle}. 
In the case where the spin-exchange interaction is ferromagnetic, the 
BEC is spin-polarized even in the absence of an external magnetic field.
Even when the spin-exchange interaction is 
antiferromagnetic, it may be also spin-polarized provided that the 
Zeeman energy is larger than the spin-exchange energy.
Accordingly, each of the weak field seeking state 
and the strong field seeking state has the same order
parameter as in $^3$He-A.

In contrast with $^3$He-A, the 
local order parameter configuration,
known as a {\it texture} in $^3$He\ \cite{woelfle}, of a spin-polarized BEC
may be easily controllable by a magnetic field. Making use of this
property, a BEC in a vortex state or with a persistent current
can be continuously created from a BEC without circulation by adiabatically
changing an external magnetic field as shown below.
The general theoretical framework for describing a spinor BEC\ \cite{stengerNature}
was given by Ohmi and Machida\cite{OhmiMachida} and independently
by Ho\cite{HoSpinor}, which turned out to be equivalent.
This framework is based on the Bogoliubov theory extended to a 
vectorial order parameter with three components,
corresponding to $m_{F}=1,0,-1$ of the $F=1$ atomic 
hyperfine state.
As a result, the generalized Gross-Pitaevskii (GP) equation has been
constructed.  They calculated low-lying collective modes
such as sound wave, spin wave, and their coupled modes
and predicted various topological defects or spin textures.

This paper is organized as follows.
In the next section, the order parameter of a BEC with $F=1$ is discussed.
We employ two sets of basis vectors and
their transformations are also considered.
Then the generalized Gross-Pitaevskii equation is introduced.
In Section III, we consider the cross disgyration
texture which is expected to appear in an Ioffe-Pritchard trap.
Section IV is the main part of the present paper.
We first consider an axially symmetric BEC 
without a current confined in an Ioffe-Pritchard trap with an optical plug.
A strong magnetic field is applied along the axis of the BEC.
Then the sign  of this axial magnetic field is
adiabatically changed so that the local 
magnetization vector flips in the end of this process.
Then it is shown that a persistent current with two units of 
circulation is created.
If the optical plug may be turned off at this stage,
we are left with a vortex line.
The topological justification of this behavior is also given. 
Observational consequences of the existence of a quantized 
vortex or a persistent current are discussed in Section V, 
where the relaxation of the order parameter after the
confining fields are turned off is studied.
Section VI is devoted to summary and discussions.

\section{Spinor Bose-Einstein Condensate}

\subsection{Spinor order parameter}

Let $F_{\alpha}\ (\alpha = x, y, z)$ be
the angular momentum operators with $F=1$.
The eigenvalues of $F_z$ are $1, 0, -1$ and their corresponding eigenvectors,
that satisfy $F_z |i \ket = i | i \ket$, are
\begin{equation}
   |1 \ket = \left(
   \begin{array}{c}
      1\\
      0\\
      0
   \end{array} \right),
   \ |0 \ket = \left(
   \begin{array}{c}
      0\\
      1\\
      0
   \end{array} \right),
   \ |-1 \ket =
   \left( \begin{array}{c}
      0\\
      0\\
      1
   \end{array} \right).
\end{equation}
In this basis, called the ZQ basis, $F_{\alpha}$ are represented as
\begin{eqnarray}
F_{x} &=& \frac{1}{\sqrt{2}} \left(
   \begin{array}{ccc}
      0 & 1 & 0 \\
      1 & 0 & 1 \\
      0 & 1 & 0 
   \end{array}
\right),
\nn\\
F_{y} &=& \frac{i}{\sqrt{2}} \left(
   \begin{array}{ccc}
      0 & -1 &  0 \\
      1 &  0 & -1 \\
      0 &  1 &  0 
   \end{array}
\right),           \label{eq:fxfyfz}
\\
F_{z} &=&  \left(
   \begin{array}{ccc}
      1 & 0 &  0 \\
      0 & 0 &  0 \\
      0 & 0 & -1 
   \end{array}
\right),
\nn
\end{eqnarray}
which satisfy the commutation relation
$[F_\alpha, F_\beta] = i F_\gamma \varepsilon_{\alpha\beta\gamma}$.
The order parameter $| \Psi \ket$ is expanded in terms of $|i \ket$
as
\begin{equation}
   |\Psi\ket = \sum_{i= 0, \pm 1} \Psi_i | i \ket .
\end{equation}
%
%

It is convenient for our purposes to introduce another set of basis vectors
$|\alpha \ket\ (\alpha = x, y, z)$
called the $XYZ$ basis which satisfy
$F_{\alpha}|\alpha \ket = 0$.
Note that $|z\ket = | 0 \ket$ by definition.
The vectors $|x \ket$ and $|y \ket$ are obtained by rotating $|z\ket$ 
around the $y$ axis and the $x$ axis by $\pm \pi/2$;
\begin{eqnarray}
   |x\ket
       &=& \exp\left(-i \frac{\pi}{2} F_y\right) |z\ket
        = \frac{1}{\sqrt{2}}(-|1 \ket + | -1 \ket),
\label{eq:xket}\\
   |y\ket
       &=& \exp\left(i \frac{\pi}{2} F_x\right) |z\ket
       =\frac{i}{\sqrt{2}}(|1\ket + |-1 \ket).
\label{eq:yket}
\end{eqnarray}
Then the order parameter $|\Psi\ket$
may be decomposed in terms of $|\alpha \ket$ as
\begin{equation}
   | \Psi\ket = \sum_{\alpha = x,y,z} \Psi_\alpha | \alpha \ket.
\end{equation}
The components $\Psi_i$ and $\Psi_{\alpha}$ are related with each other as
\begin{equation}
   \left(
      \begin{array}{c}
         \Psi_{1} \\ \Psi_{0} \\ \Psi_{-1}
      \end{array}
   \right)
   = 
   \left(
      \begin{array}{ccc}
         \frac{-1}{\sqrt{2}} & \frac{i}{\sqrt{2}} & 0
      \\
         0                   & 0 & 1
      \\
         \frac{1}{\sqrt{2}}  & \frac{i}{\sqrt{2}} & 0
      \end{array}
   \right)
   \left(
      \begin{array}{c}
         \Psi_{x} \\ \Psi_{y} \\ \Psi_{z}
      \end{array}
   \right).
\end{equation}
%
%

In a spin-polarized BEC, the weak or strong field seeking
state is represented by an order parameter:
\begin{equation}
   {\bf \Psi} = \frac{\psi}{\sqrt{2}}e^{i \alpha} (\uniM + i  \uniN)
\end{equation}
in the $XYZ$ basis, where
$\psi = \sqrt{\sum_k |\Psi_k|^2}$ and $\uniM$ and $\uniN$
are real unit vectors.
We also define
\begin{equation}
   \uniL = \uniM \times \uniN,
\end{equation}
which represents the direction of the atomic hyperspin.
The three real vectors $( \uniM , \uniN,  \uniL )$ form a triad.

In the above explanation the direction of the axis of quantization is named $z$.
We may take this direction arbitrary.
When the axis is parallel to the direction of
the magnetic field (B in this paper),
the Zeeman energy term is written most simply.
We call this $B$-quantized (BQ) notation.
When the direction of the axis does not vary spatially
and parallel to the $z$ axis, we call this $z$-quantized (ZQ) notation.
The kinetic energy term is written simply in this way.
The numerical details are given in Appendix\ \ref{sec:Bmatrix}.

\subsection{Gross-Pitaevskii equations}

The time-dependent form of the Gross-Pitaevskii
(GP) equation with a spin-degree of freedom obtained
by Ohmi and Machida\cite{OhmiMachida}, originally in the $XYZ$ basis,
can be rewritten in the ZQ notation as
\begin{eqnarray}
   i{\partial\over\partial t}\Psi_{j}
   &=&
   \{
      h_{jk} + g_n \delta_{jk} \sum_l |\Psi_l|^2
\nn\\&&
      +g_s \sum_{\alpha} \sum_{lp} \left(
         \Psi_{l} (F_{\alpha})_{lp} \Psi_{p}
      \right) (F_{\alpha})_{jk}
   \} \Psi_{k}
\label{eq:gptime}
\end{eqnarray}
where 
\begin{eqnarray}
   h_{jk}({\bf r})&=&
   \left(
      -\frac{\hbar^2\nabla^2}{2m} -\mu
      + V({\bf r})
   \right)\delta_{jk} - {\mathcal{B}}_{jk},
\label{eq:HamilOne}\\
   {\mathcal{B}} &=&
   \left(\begin{array}{ccc}
      B_z     & \frac{B_x - iB_y}{\sqrt{2}} & 0
   \\
      \frac{B_x + iB_y}{\sqrt{2}} & 0       & \frac{B_x - iB_y}{\sqrt{2}}
   \\
      0       &  \frac{B_x + iB_y}{\sqrt{2}} & -B_z   
\label{eq:omegaMatrix}
   \end{array}\right),
\end{eqnarray}
$m$ is the mass of an atom and $\bv{B}=(B_x, B_y, B_z)
$ is a magnetic field scaled so that the amplitude is the 
Zeeman energy of the atom.
The potential $V(\bf{ r})$ is spin-independent and
$i,j=0,\pm 1$ are the spin indices in the ZQ basis.
The parameters $g_n$ and $g_s$ denote the 
strength of the interactions.
The relationship between $(B_x, B_y, B_z)$ and ${\mathcal{B}}$ is
explained in Appendix\ \ref{sec:Bmatrix}.
Time-independent solutions of GP equation are obtained by solving
\begin{eqnarray}
   0 &=&
   \{
      h_{jk} + g_n \delta_{jk} \sum_l |\Psi_l|^2
\nn\\&&
      +g_s \sum_{\alpha} \sum_{lp} \left(
         \Psi_{l} (F_{\alpha})_{lp} \Psi_{p}
      \right) (F_{\alpha})_{jk}
   \} \Psi_{k}. \label{eq:gpstat}
\end{eqnarray}
The above equations are derived from the Hamiltonian
\begin
{eqnarray}
  H
  &=&
  \int \!\!  d{\bv r} \sum_j
  \Psi^{\dagger}_j({\bv r}) h_{jk}({\bv r}) \Psi_k({\bv r})       \nonumber
\\&&
  +\frac{g_n}{2} \sum_{jk} \Psi_j^{\dagger} ({\bv r})
  \Psi_k^{\dagger}({\bv r})
  \Psi_k({\bv r})
  \Psi_j({\bv r})                   \nonumber
\\&&
  +\frac{g_s}{2} \sum_{\alpha}
  \Biggl(
    \sum_{jk}
    \Psi_j^{\dagger}({\bv r}) (F_{\alpha})_{jk} \Psi_k({\bv r})
  \Biggr)^2.  \label{eq:Hamiltonian}
\end{eqnarray}
%
%

\section{Strong and weak field seeking states in Ioffe-Pritchard trap}
\label{purequad}

We consider a system of BEC which is uniform along the $z$ axis.
The cylindrical coordinates ${\bv r} = (r, \phi, z)$ 
are introduced.
Suppose that an Ioffe-Pritchard field
\begin{equation}
   {\bv B} = (B_{\perp}(r) \cos \phi , - B_{\perp}(r) \sin \phi, B_z)
\label{eq:case0jiba}
\end{equation}
is applied to the system.
We treat two-dimensional system (uniform along the $z$ axis)
in the following calculations and $B_z$ is treated as a constant,
which differs from the usual Ioffe-Pritchard field.
There should be a blue-detuned laser beam penetrating along the $z$ axis to
prevent the atoms from escaping from the trap by spin-flipping.
Ohmi and Machida \cite{OhmiMachida} have shown that
there appears the cross disgyration when $B_z=0$.

Let us derive the configuration of the condensate in 
this system.
The ${\mathcal{B}}$-matrix Eq.\ (\ref{eq:omegaMatrix}) becomes
\begin{equation}
   \left(\begin{array}{ccc}
	B_z
		& B_{\perp} \frac{e^{i\phi}}{\sqrt{2}}

			& 0
   \\
	B_{\perp} \frac{e^{-i \phi}}{\sqrt{2}}
		& 0
          
			& B_{\perp} \frac{e^{i \phi}}{\sqrt{2}}
   \\

	0
		& B_{\perp} \frac{e^{-i \phi}}{\sqrt{2}}
			& -B_z
   \end{array}\right).
\label{eq:bmatrix}
\end{equation}
The eigenvalues of ${\mathcal{B}}$ are $\pm B$ and $0$
where $B=\sqrt{B_{\perp}^2+ B_z^2}$, and the corresponding 
eigenvectors, denoted by $| i \ket_{\rm BQ}$, are
\begin{eqnarray}
    |\pm 1 \ket_{\rm BQ}
    &=&
    \frac{1}{2 B} \left(
	\begin{array}{c}
	    (B \pm B_z) e^{i \phi}
	\\
	    \pm \sqrt{2} B_{\perp}
	\\
	    (B \mp B_z) e^{- i \phi}
	\end{array}
    \right),
\\
    | 0 \ket_{\rm BQ}
    &=& 
    \frac{1}{\sqrt{2} B } \left(
	\begin{array}{c}
		- B_{\perp} e^{i \phi}
	\\
		\sqrt{2} B_{z}
	\\
		B_{\perp} e^{- i \phi}
	\end{array}
    \right).
\end{eqnarray}
The vectors $|1 \ket_{\rm BQ}$ and $|-1 \ket_{\rm BQ}$ are identified with
the strong field seeking state and the weak field seeking 
state respectively.
Accordingly when the whole system is in the strong or the 
weak field seeking state, the order parameter is written
in terms of these vectors as
$|\Psi \ket = \Psi(\bv{r}) |\pm 1 \ket_{\rm BQ}$.
The ${\mathcal{B}}\Psi$ term in Eqs.\ (\ref{eq:gptime}) and (\ref{eq:HamilOne})
is then 
simplified to $\pm B(r) \Psi$ so that the GP equation takes the form
\begin{equation}
    i \frac{\partial \Psi}{\partial t}
    = 
    \left\{
	-\frac{\hbar^2\nabla^2}{2m} - \mu + V(r)
	\mp B(r) + (g_n + g_s) |\Psi|^2
    \right\} \Psi.
\label{eq:gp-weak-strong}
\end{equation}
%
%
We have ignored the small corrections,
which comes from the spatial derivative,
to the kinetic energy term.
This is the usual GP equation without the
spin degrees of freedom.

When $B_z = 0$, the order 
parameter with the highest eigenvalue
corresponds to the 
strong field seeking state is
\begin{equation}
   \left(
      \begin{array}{c}
         \Psi_{1} \\ \Psi_{0} \\ \Psi_{-1}
      \end{array}
   \right)
   = 
   \psi
   \left(
      \begin{array}{c}
         \frac{1}{2}e^{i \phi}
      \\
         \frac{1}{\sqrt{2}}    
      \\

         \frac{1}{2}e^{- i \phi}
      \end{array}
   \right)
   e^{iw\phi},
\label{eq:psicross}
\end{equation}
where $w$ is an integer
and $\psi$ is the amplitude of ${\bf \Psi}$.
In the $XYZ$ basis, this is rewritten as
\begin{equation}
   \left(
      \begin{array}{c}
         \Psi_{x} \\ \Psi_{y} \\ \Psi_{z}
      \end{array}
   \right)
   = 
   \psi
   \left(
      \begin{array}{c}
         - i \frac{1}{\sqrt{2}} \sin \phi
      \\
         - i \frac{1}{\sqrt{2}} \cos \phi
      \\
         \frac{1}{\sqrt{2}}
      \end{array}
   \right)
   e^{iw\phi}.
\label{eq:psicross-xyz}
\end{equation}
The corresponding $\uniM, \uniN$ and $\uniL$ vectors are
\begin{eqnarray}
   \uniM &=&
	(\sin\phi \sin(w\phi),
		\cos\phi \sin(w\phi), 
			\cos(w\phi)),
\nn\\
   \uniN &=&
	(- \sin\phi\cos(w \phi),
		- \cos\phi\cos(w \phi),
			\sin(w \phi)),
\label{eq:MNLcross}\\
   \uniL	&=& (\cos \phi, - \sin \phi, 0).
\nn
\end{eqnarray}
%
%

In the weaking field seeking state, the order parameter is written as
\begin{equation}
   \left(
      \begin{array}{c}
         \Psi_{1} \\ \Psi_{0} \\ \Psi_{-1}
      \end{array}
   \right)
   = 
   \psi
   \left(
      \begin{array}{c}
         \frac{1}{2}e^{i \phi}

      \\
         - \frac{1}{\sqrt{2}}    
      \\

         \frac{1}{2}e^{- i \phi}
      \end{array}
   \right)
   e^{iw\phi}
\label{eq:psicross-weak}
\end{equation}
or
\begin{equation}
   \left(
      \begin{array}{c}
         \Psi_{x} \\ \Psi_{y} \\ \Psi_{z}
      \end{array}
   \right)
   = 
   \psi
   \left(
      \begin{array}{c}
         - i \frac{1}{\sqrt{2}} \sin \phi
      \\
         - i \frac{1}{\sqrt{2}} \cos \phi
      \\
         - \frac{1}{\sqrt{2}}
      \end{array}
   \right) e^{iw\phi}.
\label{eq:psicross-xyz-weak}
\end{equation}
The corresponding triad is
\begin{eqnarray}
   \uniM &=&
	(\sin\phi \sin(w\phi),
		\cos\phi \sin(w\phi),
			-\cos(w\phi)),
\nn\\
   \uniN &=&
	(- \sin\phi\cos(w \phi),
		- \cos\phi\cos(w \phi),
			-\sin(w \phi)),
\label{eq:MNLcross-weak}\\
   \uniL	&=& (- \cos \phi, \sin \phi, 0).
\nn
\end{eqnarray}
This $\uniL$ vector field
in both (\ref{eq:MNLcross}) and (\ref{eq:MNLcross-weak})
clearly represents the cross disgyration.

\section{Creation of vortex}
In the present Section, we propose a simple method to create a 
persistent current (and also a vortex state) in a 
torus-shaped BEC from a state with no persistent current.
Then it will be shown that this persistent current is easily
transformed to a vortex.

In the following we consider two cases separately.
In Case I, the condensate is confined optically and
the spin of each atom points the direction of the magnetic field.
Namely, atoms are strong field seekers.
In Case II, the magnetic field is also used to confine the condensate.
The spin of atoms points are antiparallel to
the magnetic field and the atoms are weak field seekers.
The former case has an advantage in its theoretical simplicity.
On the other hand, the latter case 
does not require apparatus for optical confinement
except for the repulsive plug around $r=0$
and can be realizable more easily.

In the following discussions we consider a BEC of Na atoms with $F=1$.
The mass of the atom $m = 3.81 \times 10^{-26}\text{kg}$,
the interaction parameter $g_n = 4\pi \hbar ^2 a / m$, and
the scattering length $a = 2.75 \times 10^{-9} \text{m}$
are employed. We ignore the other interaction parameter $g_s$.
This is possible since the whole condensate is assumed to be
in either strong or weak field seeking state as a whole in the following. 
In those states the interaction terms in the GP equation are
reduced as shown in Eq.\ (\ref{eq:gp-weak-strong}).

The particle density is taken to be around $10^{19} \text{m}^{-3}$ and
the detailed density profile is given in each figure.
The time span of the persistent current creation process $T = 30 \text{ms}$
is chosen since this is between (the Larmor frequency)$^{-1}$
$\sim 1\text{m sec}$ 
and the life time of the condensate $\sim 1 \ \text{sec}$.
%
%

\subsection{Case I: Optical confinement}\label{case1}

The external magnetic field ${\bv B}(r, \phi, z)$ takes the form
\begin{eqnarray}
   (B_x, B_y)	&=& B_{\perp} (\cos \phi, - \sin \phi),
\nn\\
   B_z	&=& B_{z0}           \cos (\pi(1- t/T)),
\label{eq:case1jiba}\\
   B_{\perp} &=& B^{\prime}_{\perp} r \sin (\pi(1- t/T)),
\nn
\end{eqnarray}
where $t$ is the time.
The factors of the field are taken to be
$B^{\prime}_{\perp} = 200 h \ [\text{J}/\mu \text{m}]$
and
$B_{z0} = 2 \times 10^3 h\ [\text{J}]$ (note that the
scaled magnetic field represents the Zeeman energy in fact).
One finds from Eq.\ (\ref{eq:case1jiba}) that $B_z$ flips 
from $- B_{z0}$ to  $B_{z0}$ so that $\uniL$ also flips in 
the end of the evolution.
The Larmor frequency is $\omega_L \sim 0.6 \times 10^3 [\text{Hz}]$
for $B \sim 2 \times 10^3 h [\text{J}]$,
Thus $\omega_L \times T \sim 18$.
The spin-independent potential is
\begin{equation}
   V(r) = \frac{m(2\pi \nu)^2}{2}r^2
   + U \exp \left( - \frac{r^2} {2r_0^2} \right),
\end{equation}
with $\nu = 200 [\text {Hz}]$,
$U = 1 \times 10^4 h[\text{J}]$ and 
$r_0 = 5 [\mu\text{m}]$. The first term of Eq.(27) is the confining
potential while the second term is the potential produced 
by the optical plug.

\begin{figure}
   \begin{center} \leavevmode
   \epsfxsize=8cm \ \epsfbox{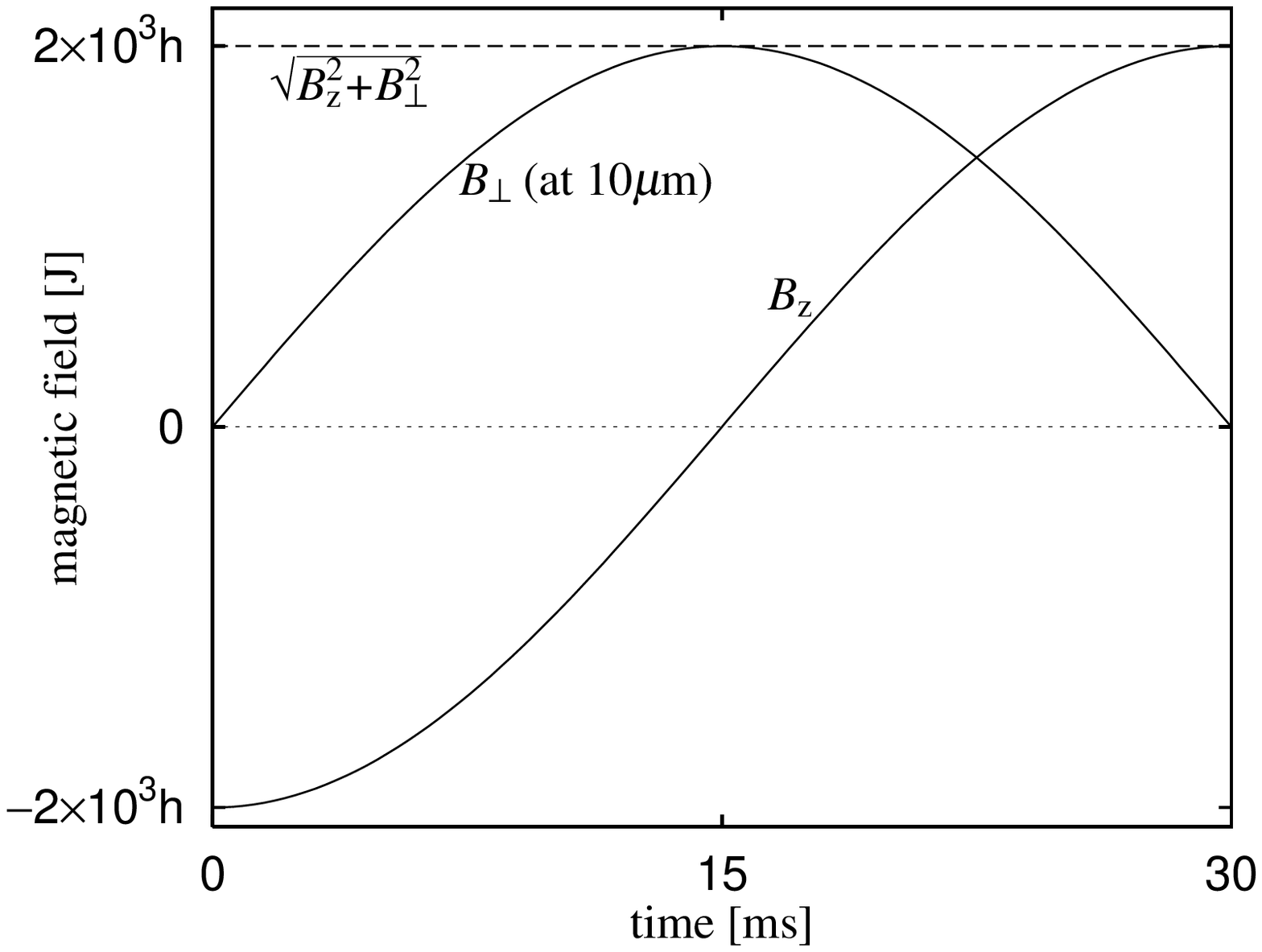} \\
   (a)\\

   \epsfxsize=8cm \ \epsfbox{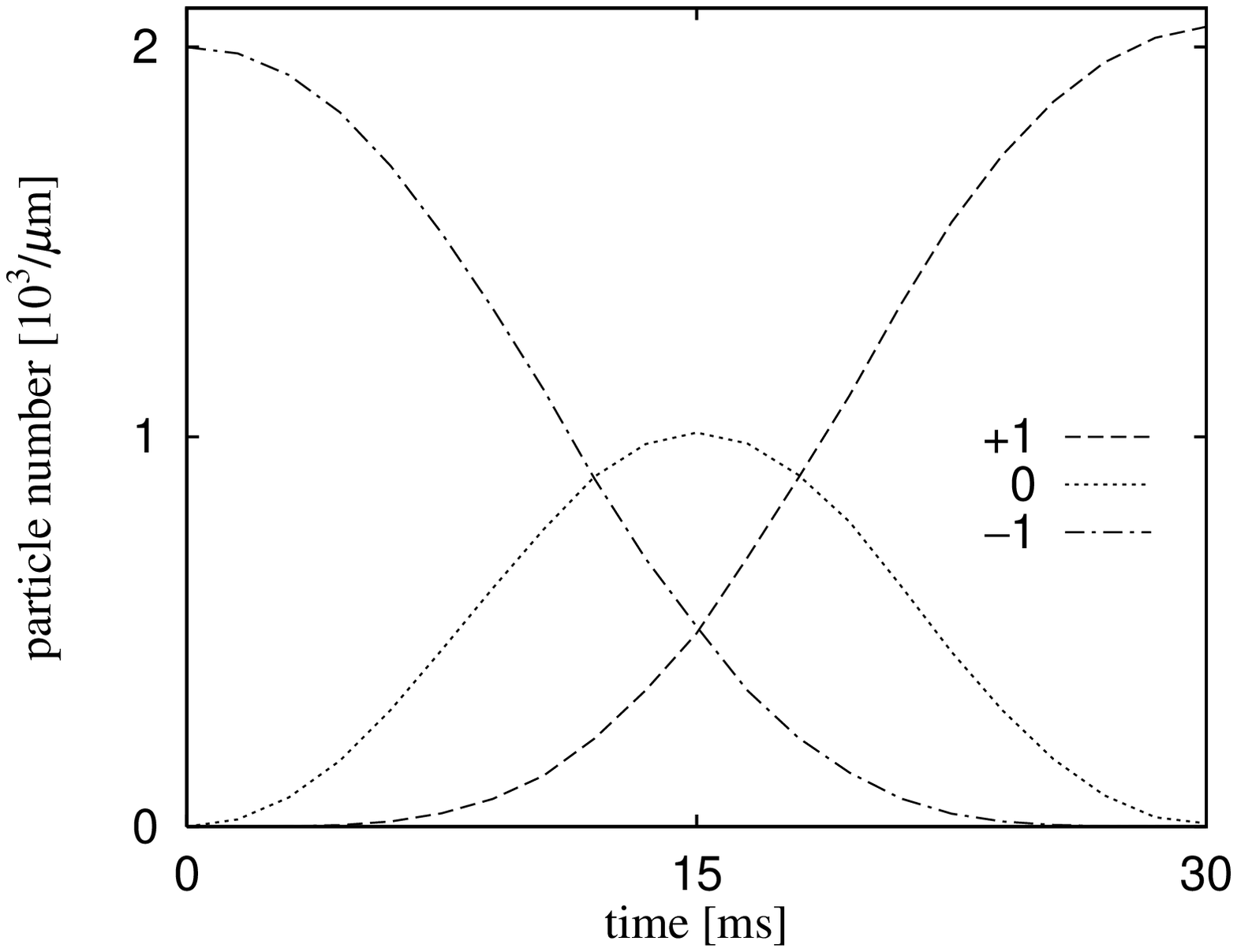} \\
   (b)\\

   \epsfxsize=8cm \ \epsfbox{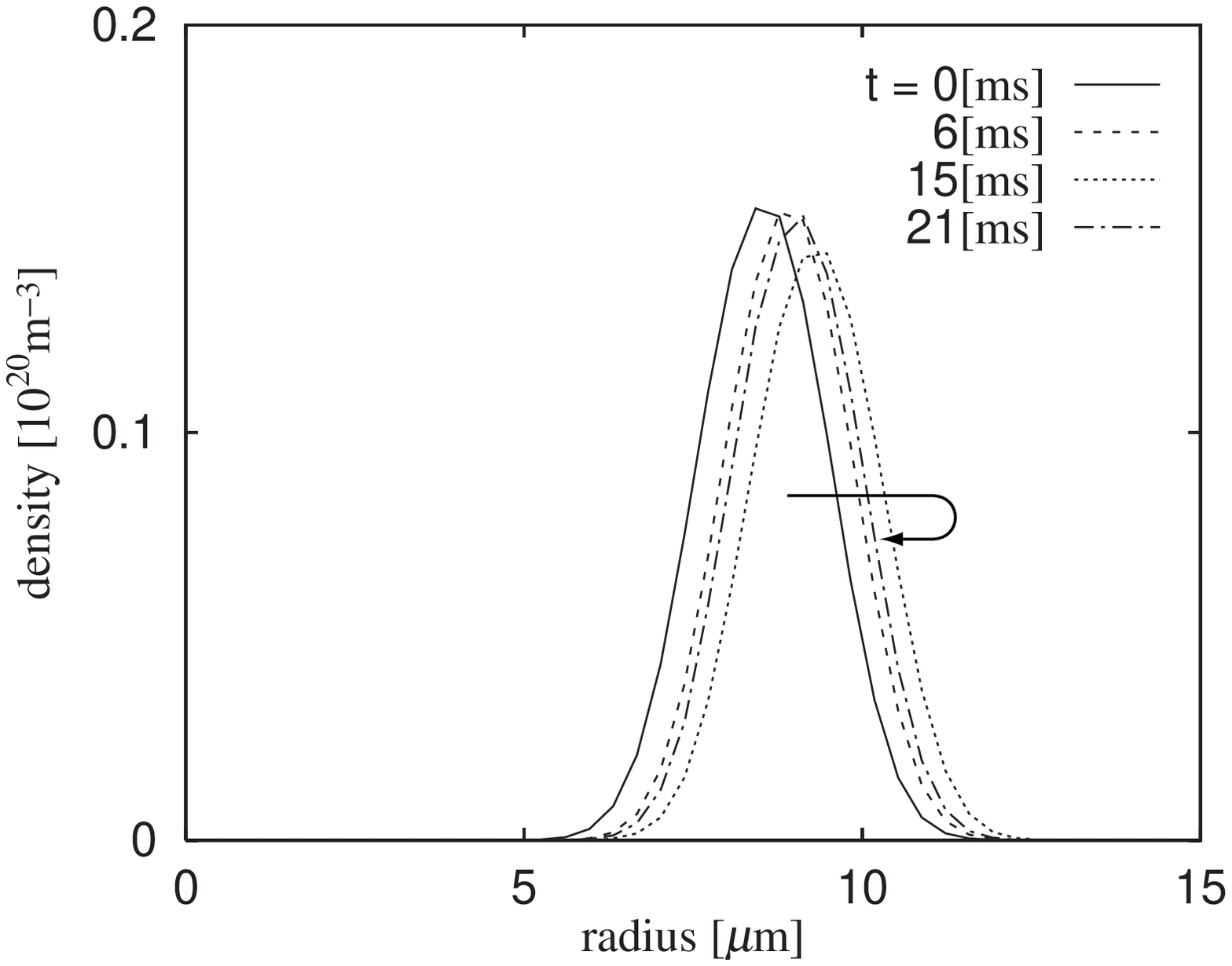} \\
   (c)\\

   \end{center}
   \caption{
   The process of the persistent current creation in Case I.
   (a) Magnetic field at $r=10\mu\text{m}$ as a function of time.
   Because $B_{\perp} \propto r$, the total magnetic field varies
   slightly at $r \ne 10\mu\text{m}$.
   (b) Particle numbers $N_k$ as a function of time.  
   The condensate has $\Psi_{-1}$ component only at $t=0$ and 
   $\Psi_{1}$ component only at $t=T$.
   (c) Total number density distribution.
   The condensate is almost fixed at around $r=9 \mu\text{m}$ by
   the optical potential $V(r)$. However
   the change of the total magnetic
   field [see caption (a)] causes the change in the radial distribution 
   as shown here.
   }
   \label{fig:case1jiba}
\end{figure}

We obtained the order parameter profile by numerical
integration of the time dependent GP equation (\ref{eq:gptime}).
The initial state is taken to be the ground state with no circulation.
The magnetic field changes the direction slowly from upward to
downward according to Eq.\ (\ref{eq:case1jiba}) as shown 
in Fig.\ \ref{fig:case1jiba}(a) so that the atoms remain in
the strong field seeking state.
The change in the number of the $k$-th component
\begin{equation}
   N_k(t) = \int 
|\Psi_k({\bf r}, t)|^2 d^2{\bf r}\quad (k=-1, 0, 1)
\end{equation}
is shown in Fig.\ \ref{fig:case1jiba}(b).
The total particle density
\begin{equation}
   n(r,t) = \sum_k |\Psi_k(r, t)|^2
\end{equation}
changes with the 
magnetic field as shown in Fig.\ \ref{fig:case1jiba}(c).

\begin{figure}
   \begin{center} \leavevmode
   \epsfxsize=8cm \ \epsfbox{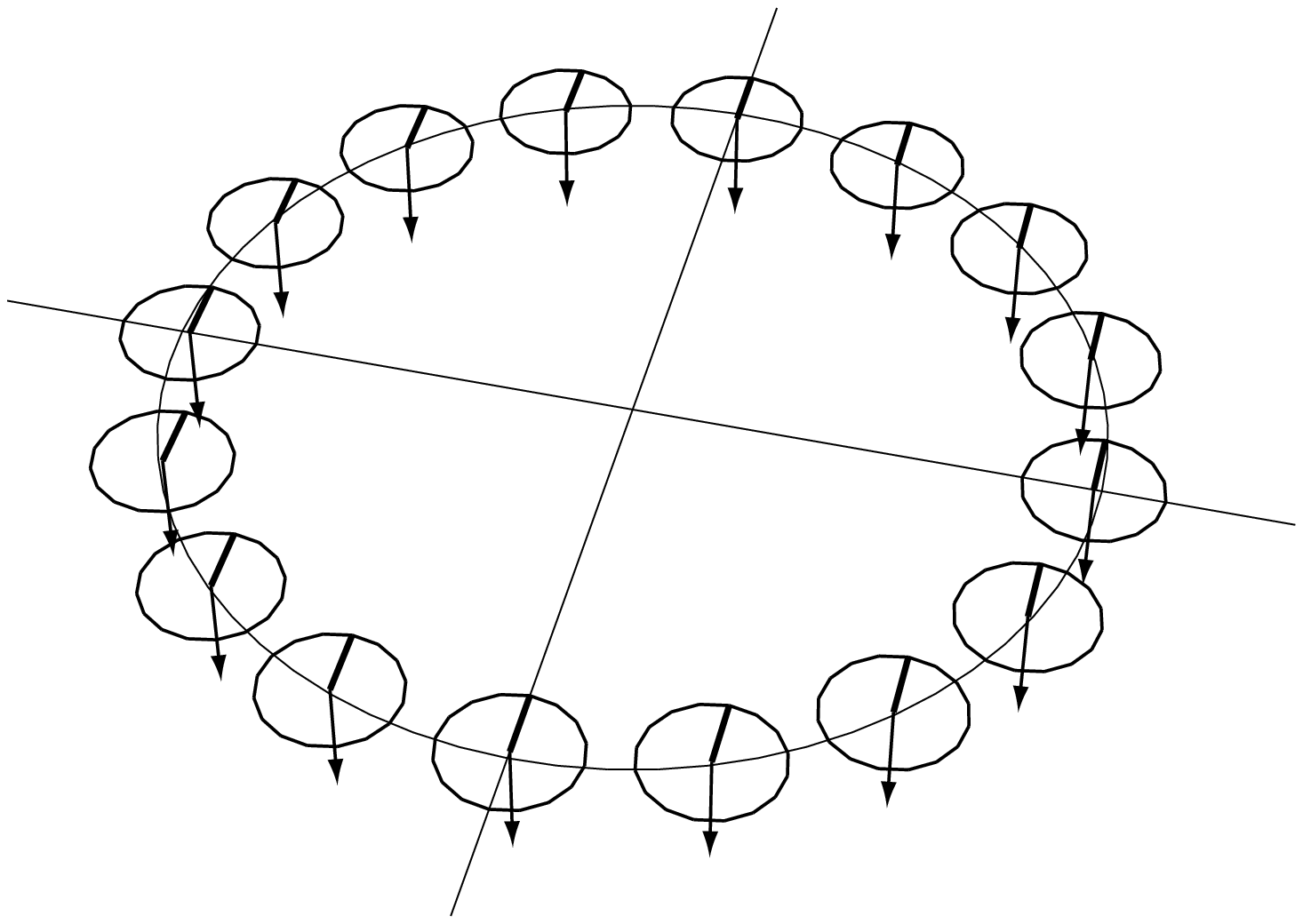} \\
   (a)\\

   \epsfxsize=8cm \ \epsfbox{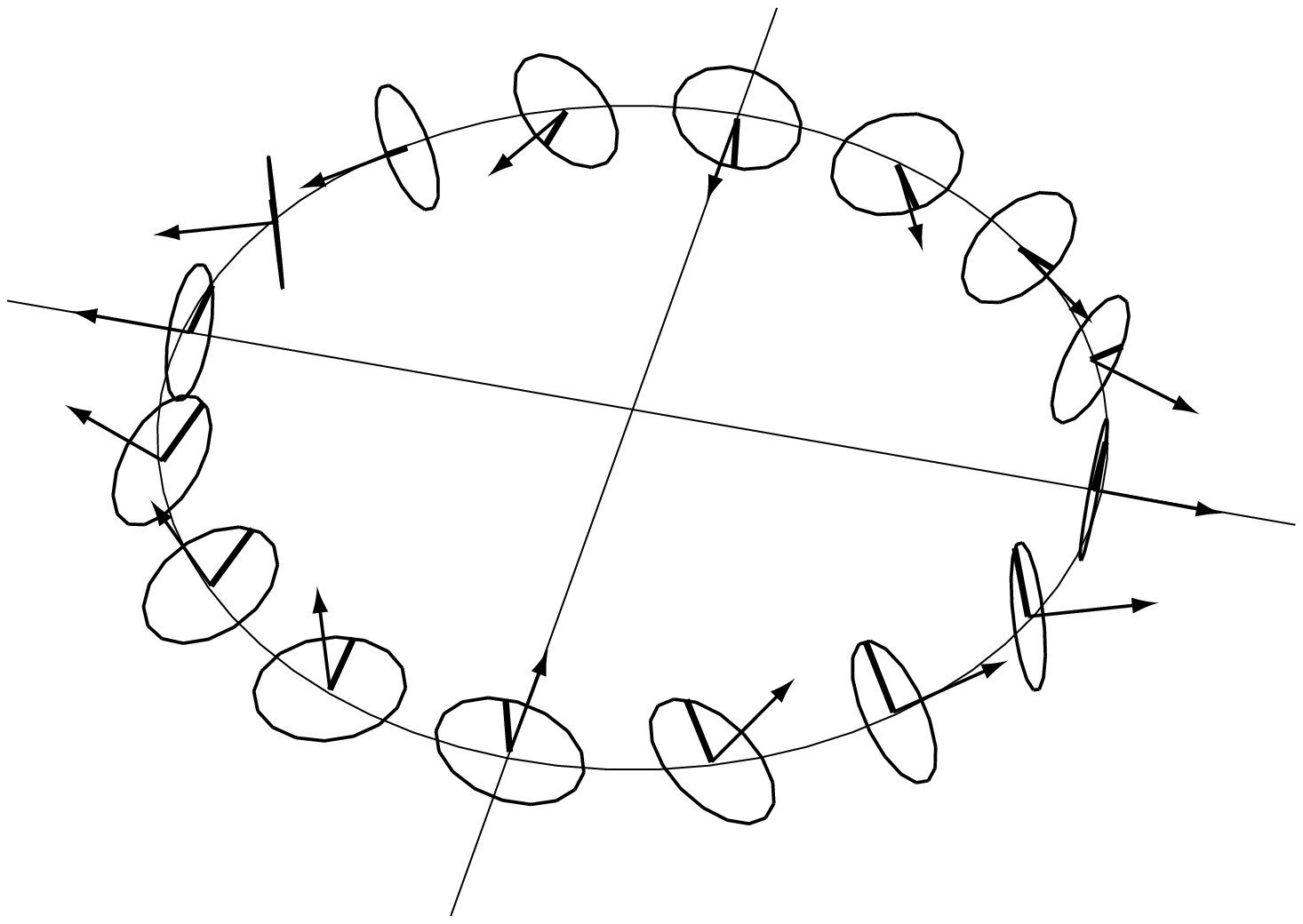} \\
   (b)\\

   \epsfxsize=8cm \ \epsfbox{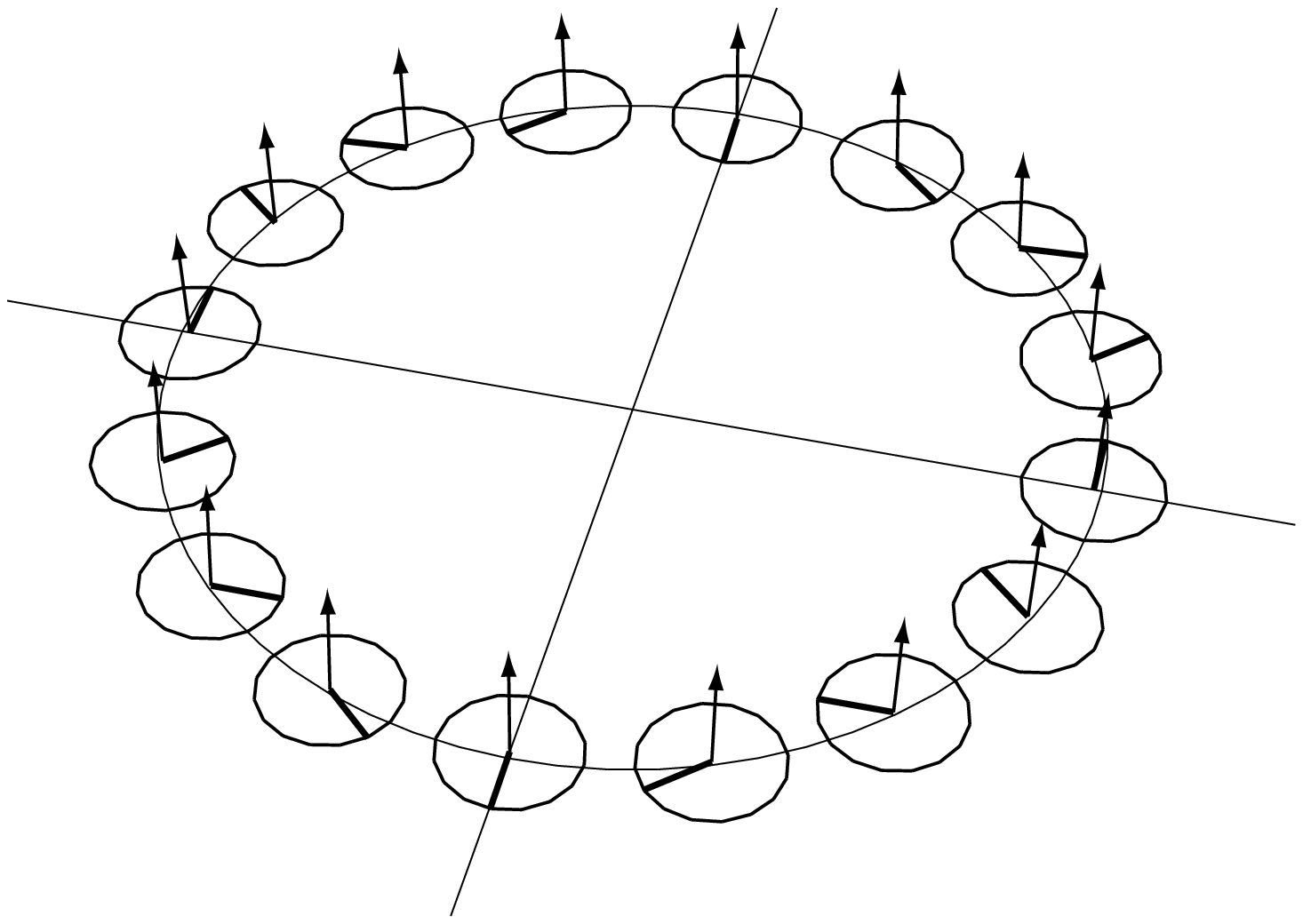} \\
   (c)\\

   \end{center}
   \caption{ 
      The triad configurations for $t = 0, 15,$ and $30\ \text{ms}$.
      The arrows denote $\uniL$ while $\uniM$ and $\uniN$ are on the
      disk.
      The line on the disk is $\uniM$.
      Note that there is no need to draw $\uniN$ 
      since it is uniquely given by $\uniL \times \uniM$.  
      When $t=0$ and $30\ \text{ms}$, $\uniL$ point down and up 
      respectively.
      When $t=15\ \text{ms}$, $\uniL$ lies almost on the $xy$ plane.
   }
   \label{fig:lmn}
\end{figure}

The resulting triad configurations are shown in Figs.\ \ref{fig:lmn}
for $t = 0, 15, 30\ [\text{ms}]$.
Figure\ \ref{fig:lmn}(a) shows the initial vector configurations.
Figure\ \ref{fig:lmn}(b)
shows the vector configurations when $t = 15\text{ms}$.
The $\uniL$-texture is nothing but the cross disgyration 
explained in the previous Section since $B_z$ vanishes now.
We see that the vectors $\uniM$ and $\uniN$
rotates around
$\uniL$ by $2\pi$ as
we go around the $z$ axis once.
Finally when $t = 30\text{ms}$, we obtain a texture with $\uniL$
almost points up everywhere.
The vectors $\uniM$ and $\uniN $ rotates around $\uniL$ by $4\pi$
as one goes around the $z$ axis once in this case
and one finally obtained a 
uniform $\uniL$-texture with a circulation
with the winding number 2.

Now that a persistent current is created,
it is easy to transform this into a vortex.
The BEC has the $k=1$ component only at $t=30 \text{ms}$.
Then there are no atoms near the axis $r=0$ since the
centrifugal force prevents the atoms to come close to the axis.
Thus one may simply turn off the optical plug to obtain a vortex.
The details are analyzed in next section.


\subsection{Case II: Magnetic confinement}

\begin{figure}
   \begin{center} \leavevmode
   \epsfxsize=8cm \ \epsfbox{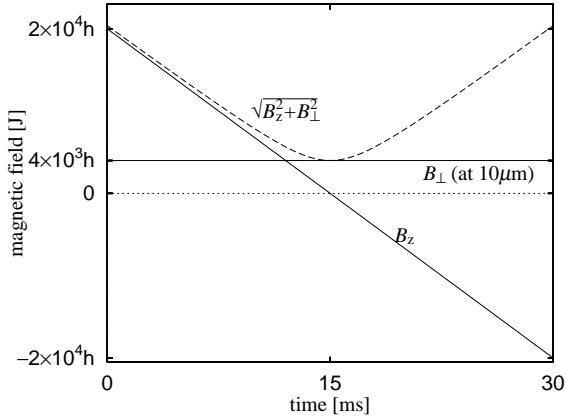} \\
   (a)

   \epsfxsize=8cm \ \epsfbox{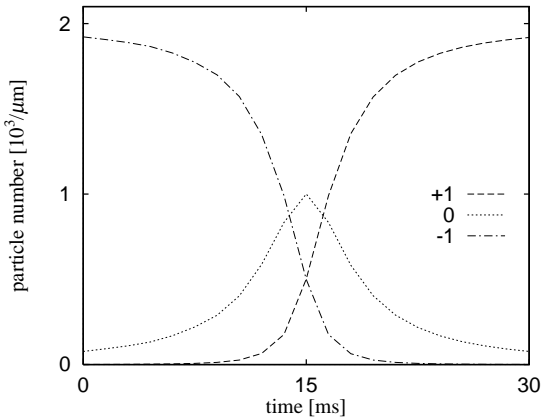} \\
   (b)

   \epsfxsize=8cm \ \epsfbox{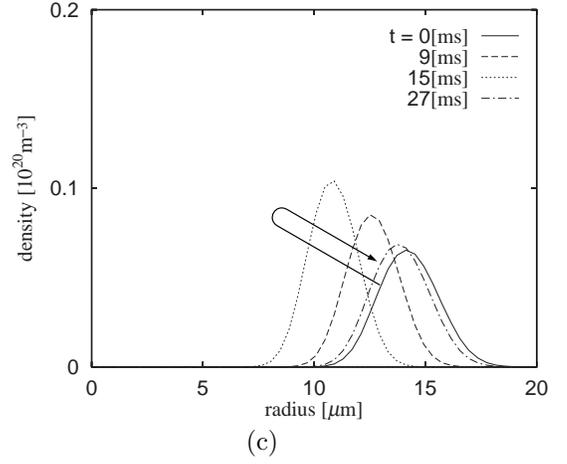} \\
   (c)

   \end{center}
   \caption{ 
   The process of the persistent current creation in Case II.
   (a) Magnetic field at $r=10\mu\text{m}$ as a function of time.
   Initially $B_z$ is positive, in contrast with Case I.
   (b) Particle numbers $N_k$ as functions of time.
   The condensate is mostly made of $\Psi_{-1}$ component at $t=0$ while 
   mostly of $\Psi_{1}$ component at $t=T$.
   (c) Total density distribution.  Contrary to Case I,
   the condensate is confined by the $B_{\perp}$ part of the magnetic field
   at larger $r$.
   Then the radial change is more eminent than that of Case I.  
   }
   \label{fig:case2jiba}
\end{figure}

In Case I, the condensate is confined with spin-independent optical trap.
Here in Case II, we consider the situation where the quadrapole
magnetic field Eq.\ (\ref{eq:case0jiba}) always exists and 
the additional field $B_z$ change from a large positive 
value to a large negative value as shown in Fig.\ \ref{fig:case2jiba}(a) :
\begin{eqnarray}
   (B_x, B_y)	&=& B_{\perp} (\cos \phi, - \sin \phi),
\nn\\
   B_z &=& B_{z0} (1 - 2t / T),
\label{eq:case2jiba}\\
   B_{\perp} &=& B^{\prime}_{\perp} r.
\nn
\end{eqnarray}
Here we take 
$B^{\prime}_{\perp} = 400 h\ [\text{J}/\mu \text{m}]$,
$B_{z0} = 2 \times 10^{4} h$
and $T = 30\ [\text{ms}]$.
Contrary to Case I,
the atoms are in the weak field seeking state and
the gradient of $B_r$ is responsible for the confinement of
the condensate.
The optical plug produces a spin-independent potential
\begin{equation}
   V(r) = U \exp\left( - \frac{r^2}{2r_0^2} \right),
\end{equation}
where we take
$U = 1 \times 10^4 h[\text{J}]$ and $r_0 = 5 [\mu\text{m}]$.

The evolution of the order parameter field is analyzed 
numerically and it was found that the order parameter 
configurations in this case is essentially the same as in Case I.
Figure\ \ref{fig:case2jiba} (b) shows the temporal 
evolution of the components $N_k(t)$
and Fig.\ \ref{fig:case2jiba}(c) shows the total density profile
at various time.

The persistent current created here may also be 
transformed into a vortex.
The BEC is mostly made of the $k=1$ component at $t=30 \text{ms}$
and there are only a small number of atoms near $r=0$.
Suppose the optical plug is turned off.
Then atoms will escape but this process should be very slow.
Thus one expects that the vortex is stable for a considerable
period of time.

%

\subsection{Mathematical analysis of continuous creation of circulation}

It may be surprising that we have {\it continuously} 
created a persistent current (a vortex) from a system 
without circulation.
Mathematically this is justified by invoking homotopy theory.
Let us denote a rotation around direction $\uniN$ by an
angle $\alpha$ by a ``vector'' $\alpha \uniE$.
This rotation is expressed as a rotation matrix
\begin{equation}
   R(\uniE, \alpha)
   =
   (1 - \cos \alpha) \hat{n}_i \hat{n}_j
   + \cos \alpha \delta_{ij}
   - \sin \alpha \varepsilon_{ijk}
   \hat{n}_k.
\end{equation}
Since $\alpha$ may be restricted within the 
region $0 \leq \alpha \leq \pi$,
the set of all the 
rotations is represented by a ball $B^3$
with the radius $\pi$.
Note however that the points $\pi \uniE$ and $-\pi \uniE$ corresponds to
equivalent rotations. Thus all the 
antipodal points on the surface of the
ball must be 
identified. This space $B^3/Z_2$ is called the 
three-dimensional
real projective space, denoted by $RP^3$.

\begin{figure}
   \begin{center} \leavevmode
   \epsfxsize=5cm \ \epsfbox{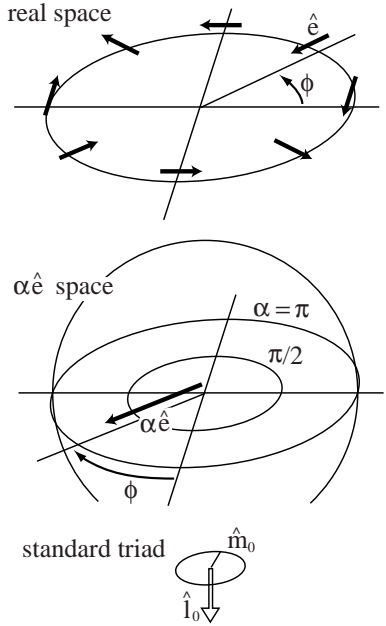} \\
   \end{center}
   \caption{The real space configuration of $\hat{\uniE}$,
   the $RP^3$ space and the standard triad.
   The thick arrows are $\alpha \uniE$ vectors.}
   \label{fig:baseLMN}
\end{figure}

Let us take a ``standard'' triad $(\uniM_0, \uniN_0, \uniL_0)$
shown in Fig.\ \ref{fig:baseLMN}.
Then an arbitrary 
triad is obtained by applying a certain
rotation $\alpha \
uniE$ to the standard triad.
Thus the local vector 
configuration is in one-to-one
correspondence with a point 
in $RP^3$.

Consider an order parameter configuration shown
in Fig.\ \ref{fig:lmn}(a).
When one circumnavigates the 
circle, one finds that all the triads along
the circle are 
obtained from the standard one by applying no rotations,

namely $\alpha = 0$.
Thus this circle is mapped to the 
origin of $RP^3$.
Next consider the triads in Fig.\ \ref{fig:lmn}(b).
As one goes along the circle, one finds 
that the standard triad is rotated
by an angle $\pi / 2$ 
around the axis
$\uniE = (-\sin \phi, -\cos \phi, 0)$ shown
in Fig.\ 4
Thus this circle is mapped to a circle in $RP^3$ 
with the radius $\pi/2$,
see Fig.\ 4.
Finally, consider
the triads in Fig.\ \ref{fig:lmn}(c).
All the $\uniL$-
vectors point up and the standard triad is rotated
by $\pi$
around the axis $\uniE$ given above. Thus the the circle
in 
Fig.\ 2(c) is mapped to a great circle with the radius $\pi$.

The change of the images in $RP^3$, namely a point $\to$
a circle with the
radius $\pi/2$ $\to$ a circle with the 
radius $\pi$, is continuous
(or more precisely homotopic), 
which shows that the deformation of
the triads is indeed 
continuous.

In the configuration (a), the vector $\uniM$
does not rotate
around $\uniL$ and therefore there is no 
current flowing around the circle.
In configuration (b), 
however, $\uniM$ rotates around $\uniL$
once while one goes
along the circle once, which implies a vortex of
winding 
number 1.
Similarly, $\uniM$ winds around $\uniL$ twice in 
the configuration (c)
and the vortex has the winding number 2.
We stress again that the creation of this winding number 
(or circulation)
is continuous and the final configuration 
is stable
so far as the external magnetic field forces $\uniL$ to point upward.

\subsection{Angular momentum analysis of current creation}

The continuous current creation can be explained more generally
using the eigenvectors of the ${\mathcal{B}}$-matrix with
the highest and lowest eigenvalues in the BQ basis.
The magnetic field expressed in ${\mathcal{B}}$-matrix
is the dominant factor to determine the behavior
of the order parameter ${\bf \Psi}$.
We first discuss the BEC with $F=1$, which we have 
analyzed in the present paper.

The magnetic fields Eq.\ (\ref{eq:case1jiba}) in Case I
and Eq.\ (\ref{eq:case2jiba}) 
in Case II are of the form Eq.\ (\ref{eq:case0jiba}).
The corresponding ${\mathcal{B}}$-matrix is given by
Eq.\ (\ref{eq:bmatrix}).
Because the magnetic fields are sufficiently strong, 
we may assume that the order parameter ${\bf \Psi}$
is proportional to the highest (upper sign) or lowest (lower sign) eigenvector
\begin{equation}
    |\pm 1 \ket_{\rm BQ}
    =
    \frac{1}{2 B} \left(
	\begin{array}{c}
	    (B \pm B_z) e^{i \phi}
	\\
	    \pm \sqrt{2} B_{\perp}
	\\
	     (B \mp B_z) e^{- i \phi}
	\end{array}
    \right).
\end{equation}
of the ${\mathcal{B}}$-matrix.
Case I uses the highest (upper sign) and Case II uses the lowest (lower sign)
eigenvector.
The order parameter is
\begin{equation}
   \left(\begin{array}{c}
      \Psi_1 \\ \Psi_0 \\ \Psi_{-1}
   \end{array}\right)
   = C
   \left(\begin{array}{l}
      (B \pm B_z) e^{i\phi}
   \\
      \pm \sqrt{2}B_{\perp}
   \\
      (B \mp B_z) e^{-i\phi}
   \end{array}\right) e^{i w \phi},
   \label{eq:orderP}
\end{equation}
where $C$ is a complex number independent of $\phi$ and $w$ is an 
integer.

Let us consider Case I. (Case II can be analyzed similarly.)
When  $t=0$, the magnetic field is $(B_{\perp}, B_z) = (0, \mp B_{z0})$.
As shown in Eq.\ (\ref{eq:orderP}),
the condensate behaves as
$\Psi_{-1} \propto e^{i (w - 1) \phi}$ and $\Psi_0 = \Psi_{1}=0$.
Because we start with a state with no circulation,
the integer $w$ must be $1$.

The components $B_z$ and $B_{\perp}$
change as shown in, for example, Fig.\ \ref{fig:case1jiba}(a)
from $t=0$ to $t=T$.
Both $B_{\perp}$ and $(B \pm B_z)$ have finite values during the 
change and the condensate ${\bf \Psi}$ stays in the strong 
field seeking state.
The phase factor $w$ will not change
during the process for the order
parameter to be defined 
uniquely. 
Thus we take $w \equiv 1$ throughout the process.
When $t=T$, the magnetic field is
$(B_{\perp}, B_z) = (0, \pm B_{z0})$.
Since $w=1$, Eq.\ (\ref{eq:orderP}) leads to 
the conclusion that
we have a condensate with 
$\Psi_{1} \propto e^{2 i \phi}$ and $\Psi_0 = \Psi_{-1}=0$.

Accordingly a vortex with the winding number 2 has been 
created.

A similar discussion is possible in the system 
with $F=2$ atoms
using the eigenvector given in Eq.\ (\ref{eq:f2vec}).
Starting from a state with no winding number,
we eventually obtain a state with the winding number 4.

\section{detection of vortex: Time of Flight imaging}

The detection of a vortex (or persistent current) has been a problem
as difficult as their creation.
We consider the relaxation of the spinor texture after the
confining field and the optical plug are turned off to 
facilitate the comparison between our theory and 
experiments, in particular the time of flight (TOF) analysis.

The temporal evolution of the BEC is described by the
time-dependent GP equation (10).
We consider Case I and Case II separately.

\subsection{Case I}

We consider three cases where the confining 
potentials are turned off
separately at
$t = 0, T/2$ and $T
\ \text{ms}$.
There is no vortex at $t=0$ while there is a 
vortex with the winding number 2 at $t=T$.
The comparison between these two cases is essential to observe our 
vortex.

\begin{figure}
   \begin{center} \leavevmode
      \epsfxsize=8cm \ \epsfbox{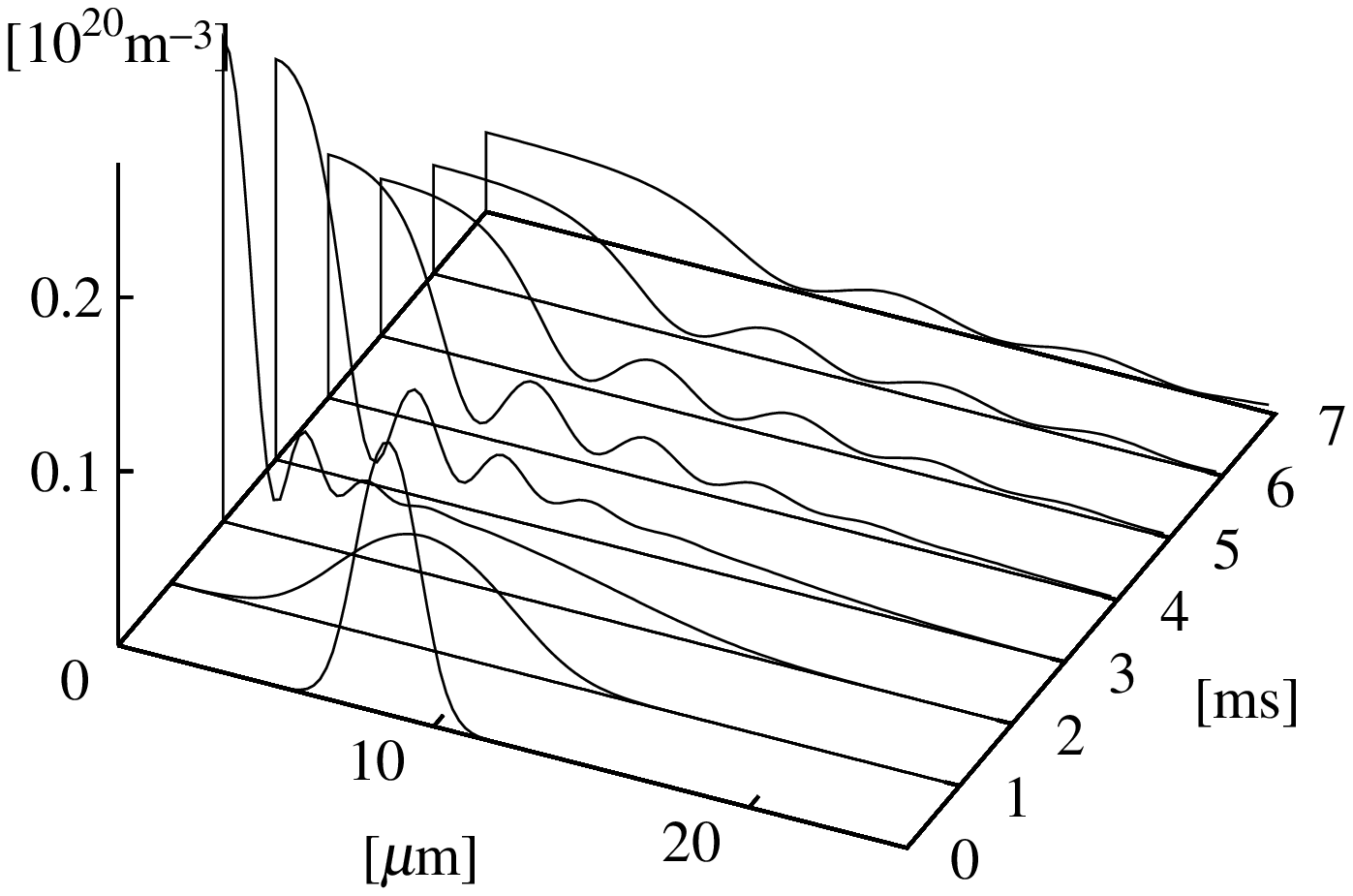} \\
      (a)

      \quad

      \epsfxsize=8cm \ \epsfbox{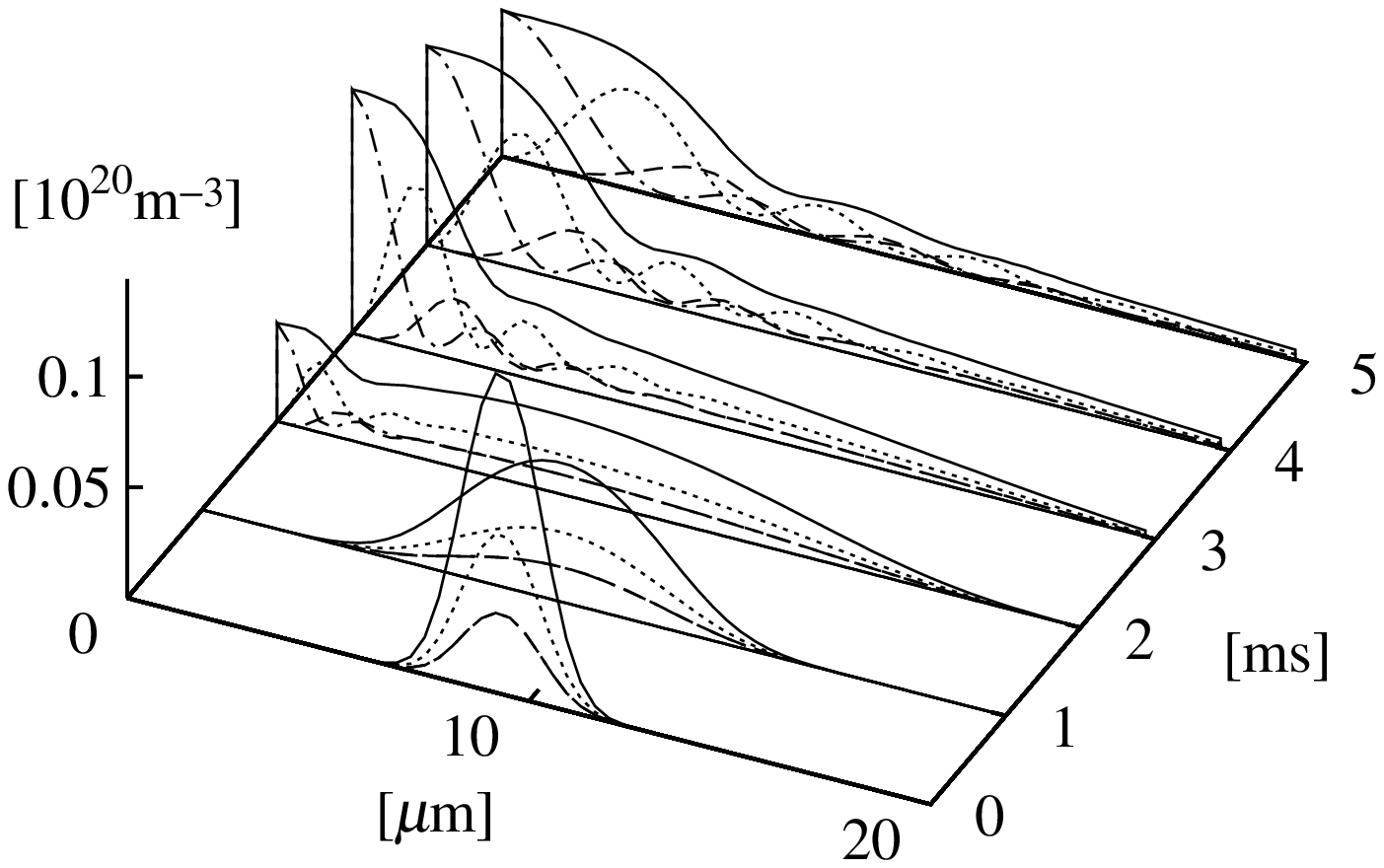} \\
      (b)
 
      \quad

      \epsfxsize=8cm \ \epsfbox{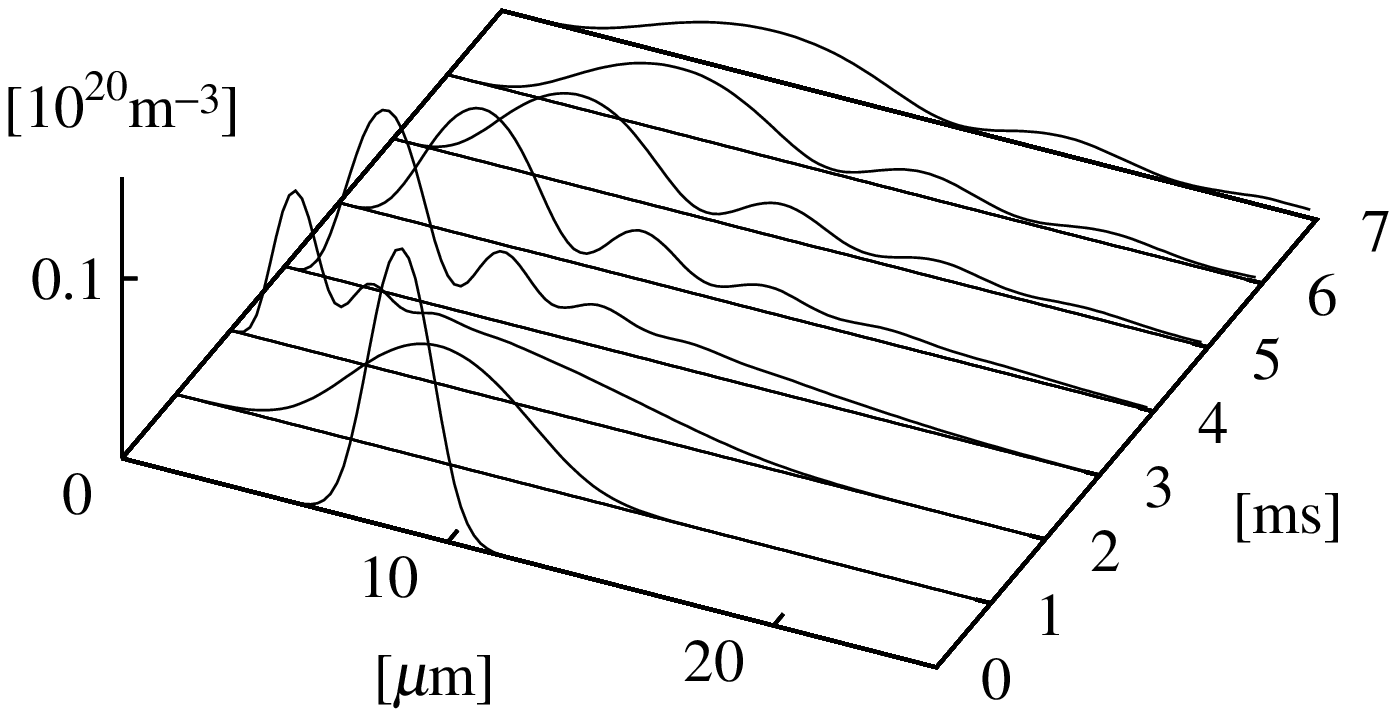} \\
      (c)
      
      \quad

   \end{center}
   \caption{Temporal evolution of the condensate.
   (a) The potentials are turned off at $t=0 $.
   The solid line is the total density purely made of
   $| \Psi_{-1}|^2$.
   (b) The potentials are turned off at $t = T/2$.
   The solid line is the total density $\sum_k
   |\Psi_k|^2$ while the dashed line, the dotted line, 
   and the dashed dotted line are components
   $|\Psi_{1}|^2, |\Psi_{0}|^2$ and $|\Psi_{-1}|^2$, respectively.
   The corresponding winding numbers are $2, 1$ and $0$.
   (c) The potentials are turned off at $t = T$.
   The solid line is
   the total number density made of $|\Psi_1|^2$ only. 
   The condensate has the winding number 2 and cannot fill 
   the central region.
   This should be compared with (a) and (b).}
   \label{fig:case1exp}
\end{figure}

(i) $t = 0$: The condensate has a component $\Psi_{-1}$ only.
The density profile at $t=0$ is determined by solving
the time-independent GP equation.
The relaxation process after the potentials are turned off is found
by solving the time-dependent GP equation Eq.\ (\ref{eq:gptime}),
whose result is shown in Fig.\ \ref{fig:case1exp}(a).
Since $\Psi_{-1}$ has no singularity at the origin, the condensate
fills the central region $(r \sim 0)$ in later time.
It is interesting to note that the components $\Psi_0$ and $\Psi_1$
do not appear in later time since the total spin must be conserved.

(ii) $t = T/2$: The cross disgyration appears in this stage.
The order parameter of this texture is given by
Eq.\ (\ref{eq:psicross}) with $w=1$.
Thus all the 
components are non-vanishing in this case.
After the potentials are turned off at $t=0$, the order parameter
relaxes as shown in Fig.\ \ref{fig:case1exp}(b).
The component $\Psi_0$ has winding number 1 while
$\Psi_{1}$ has winding number 2, and hence they cannot fill
the central region.
The central region near $r=0$ may be filled
only with $\Psi_{-1}$ component
since it has vanishing 
winding number. Note also
that the $\Psi_0$ component is 
dominant in the vicinity of $r \sim 1 \mu \text{m}$. 

(iii) $t = T$: The vector $\uniL$ points up now and
hence $\Psi _{-1} = \Psi_{0} = 0$ while $\Psi_{1} \neq 0$.
Figure\ \ref{fig:case1exp}(c) shows the temporal evolution of
the order parameter after the potentials are turned off.
Since the order parameter has a nontrivial phase factor,
it cannot fill the central region at all in later times.
Similarly to the case (i), the components $\Psi_0$ and $\Psi_{-1}$
do not appear in the relaxation process.
The absence of the condensate at $r=0$ at an arbitrary time
is a clear distinction
between the case (iii) and the rest, which may be used to show
the existence of the vortex or the persistent current experimentally.

The vacuum region around $r=0$ which shows the
existence of the vortex have length scale around $10 \mu m$
after 7ms relaxation,
and the length will be sufficient to observe experimentally.
Because the length scale of the central vacuum region is almost
same as that of the density waves at larger $r$ (outer),
the resolution of the imaging will be checked by the outer density waves.


\subsection{Case II}

\begin{figure}
   \begin{center} \leavevmode
      \epsfxsize=8cm \ \epsfbox{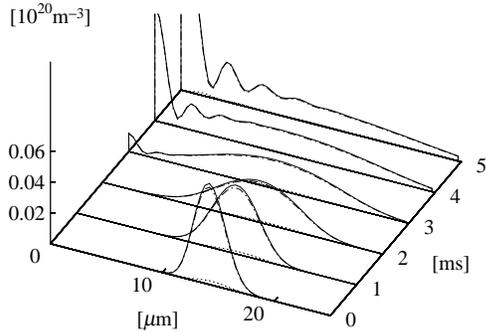} \\
      (a)

      \quad

      \epsfxsize=8cm \ \epsfbox{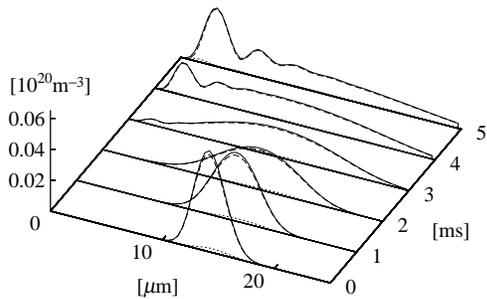} \\
      (b)
      
   \end{center}
   \caption{
      Temporal evolution of the condensates when the potentials are
      turned off at
      (a) $t = 0$ and (b) $t=T$. In both cases the description of the
      lines are the same as that in Fig.5.
   }
   \label{fig:case2exp}
\end{figure}

Figure\ \ref{fig:case2exp}(a) shows the relaxation process when the
potentials are turned off at $t=0$ while they are turned off at $t=T$
in Fig.\ \ref{fig:case2exp}(b).
Because the magnetic field is not exactly parallel or antiparallel
to the $z$ axis, the non-dominant 
component of the condensate appears slightly in both cases.

\section{summary and discussions}

In summary, we have proposed a new method to create a
persistent current and a vortex with the winding number 2
in a Bose-Einstein condensate of alkali atoms. 
The dynamics of vortex creation are simulated by solving the 
time-dependent Gross-Pitaevskii (GP) equation.
The continuity of this process is justified by invoking
homotopy theory and also by the angular momentum analysis.
The existence of the vortex may be demonstrated by 
comparing the time of flight (TOF) data
before and after the vortex creation.

It is also possible to create a 
persistent current or a vortex with the
winding number 1. 
Suppose one prepares the ground state in the
Ioffe-Pritchard field.
The resulting texture is the cross disgyration
with no winding of the $\uniM$-vector around 
the $\uniL$-vector as shown in\ \cite{OhmiMachida}.
Then apply a strong magentic field $B_z$ either parallel to
or antiparallel to the $z$ axis.
The $\uniL$-vector in the resulting texture
points up or down, depending on the 
direction of $B_z$ or whether the state is weak field 
seeking or strong field seeking.
In any case,
the $\uniM$-vector rotates around $\uniL$ once as one circumnavigates
around the $z$ axis once.
Thus one obtains a persistent current or a vortex
of the winding number 1 by simply 
preparing a sample in the Ioffe-Pritchard
trap and applying
a strong magnetic field along the $z$ axis.

The work of MN is supported partially by the
Grant-in-Aid for Scientific 
Research Fund of the Ministry of Education,
Science, Sports
and Culture, No.11640361.

\nidanOwari
\appendix

\section{magnetic field matrix} \label{sec:Bmatrix}

We consider a system of atoms with spin $F=f$.

The $F$-matrices, which are the angular momentum operators,
are given by 
\begin{eqnarray}
   F_x &=&
   \left(\begin{array}{@{}ccccc@{}}
	0 & \frac{\sqrt{2f \cdot 1}}{2}                       & & &  \\
	\frac{\sqrt{2f\cdot 1}}{2} & 0 & \frac{\sqrt{(2f-\!1)\cdot 2}}{2} & &  \\
      	& \frac{\sqrt{(2f-\!1)\cdot 2}}{2} & \ddots & \ddots      &  \\
      	& & \ddots & 0	& \frac{\sqrt{1 \cdot 2f}}{2}                \\
      	& & & \frac{\sqrt{1 \cdot 2f}}{2}                         & 0
   \end{array}\right),
\nn\\
   F_y &=& 
   \left(\begin{array}{@{}ccccc@{}}
	0	& \frac{\sqrt{2f \cdot 1}}{2i}	&	&	&\\
	-\frac{\sqrt{2f\cdot 1}}{2i}	& 0	& \frac{\sqrt{(2f-\!1)\cdot 2}}{2i} &	&\\
      	&-\frac{\sqrt{(2f-\! 1)\cdot 2}}{2i}& \ddots & \ddots &\\
      	&	& \ddots & 0	& \frac{\sqrt{1 \cdot 2f}}{2i}\\
      	&	&	& -\frac{\sqrt{1 \cdot 2f}}{2i}	& 0
   \end{array}\right)
\\
   (F_z)_{jj} &=& f + 1 - j \quad (j = 1, 2, \cdots , 2f + 1).
\nn
\end{eqnarray}
They satisfy the commutation relation
$[F_\alpha, F_\beta] = i F_\gamma \varepsilon_{\alpha\beta\gamma}$.

We write the magnetic field with the $B$ vector
\begin{equation}
   \left(\begin{array}{c}
      B_x \\ B_y \\ B_z 
   \end{array}\right)
   =
   B\left(\begin{array}{c}
      \sin \theta_y \cos \theta_z \\ \sin \theta_y \sin \theta_z  \\ \cos \theta_y
   \end{array}\right)
   =
   \left(\begin{array}{c}
      B_{\perp} \cos \theta_z \\ B_{\perp} \sin \theta_z \\ B_z
   \end{array}\right)
\end{equation}
where $0 \leq \theta_y \leq \pi$ and $0 \leq \theta_z < 2\pi$.
The amplitude $|\vec{B}|$ is scaled so that it represents the 
Zeeman energy.

The order parameter of a BEC is written 
with a vector of $2f+1$ components
and operators are expressed in terms of $(2f+1) \times (2f+1)$
square matrices.
We call the operator for the Zeeman energy ${\mathcal{B}}$-matrix.
We can choose the quantization axis so that
the ${\mathcal{B}}$-matrix is proportional to the 
matrix $F_z$.
We call this the $B$ quantized (BQ) notation 
because $\bv{B}$ is proportional to the quantization axis.

The ${\mathcal{B}}$ matrix in the ZQ notation is obtained 
by successive spatial rotations $\theta_y$ along the $y$ 
axis and $\theta_z$ along the $z$ axis as
\begin{eqnarray}
   {\mathcal{B}}_{\text{ZQ}} &=& U^{\dagger} {\mathcal{B}}_{\text{BQ}} U ,
\nn\\
   U^{\dagger} &=& \exp(- i F_z \theta_z) \exp(- i F_y \theta_y) .
\end{eqnarray}

Let us study a few examples. When $f=1$,
\begin{equation}
   {\mathcal{B}}_{\text{ZQ}} =
   \left(\begin{array}{ccc}
	B_z
		& \frac{B_{\perp}}{\sqrt{2}} e^{- i \theta_z}   
			& 0
   \\
	\frac{B_{\perp}}{\sqrt{2}}  e^{i \theta_z}
		& 0
			&  \frac{B_{\perp}}{\sqrt{2}} e^{- i \theta_z}
   \\
	0
		& \frac{B_{\perp}}{\sqrt{2}} e^{i \theta_z}   
			& -B_z
   \end{array}\right).
\end{equation}

The eigenvector with the lowest eigenvalue is
\begin{eqnarray}
   U^{\dagger} 
   \left(\begin{array}{c}
      0 \\ 0 \\ 1
   \end{array}\right)
   &=&
   \left(\begin{array}{c}
      \frac{B - B_z}{2} e^{-i \theta_z}
   \\
      - \frac{B_{\perp}}{\sqrt{2}}
   \\
      \frac{B + B_z}{2} e^{i \theta_z}
   \end{array}\right).
\end{eqnarray}
When $f=2$,
\begin{eqnarray}
   {\mathcal{B}}_{\text
{ZQ}} =
   \left(\begin{array}{ccccc}
	2 B_z
		& B_{\perp} e^{- i \theta_z}
			& 0	& 0	& 0
\\  
	B_{\perp} e^{i \theta_z}
		&   B_z
			&  \sqrt{\frac{3}{2}} B_{\perp} e^{- i \theta_z}
				& 0	& 0
\\  
	0  
		&  \sqrt{\frac{3}{2}} B_{\perp} e^{i \theta_z}
	 		& 0
				& \sqrt{\frac{3}{2}} B_{\perp} e^{- i \theta_z}
					& 0
\\
	0
		& 0
			& \sqrt{\frac{3}{2}} B_{\perp} e^{i \theta_z}
				& - B_z
					& B_{\perp} e^{- i \theta_z}
\\
	0	& 0	&   0 
				& B_{\perp} e^{i \theta_z}
					& -2 B_z
   \end{array}\right).
\end{eqnarray}
The eigenvector with the lowest eigenvalue is

\begin{equation}
   U^{\dagger} 
   \left(\begin{array}{c}
      0 \\ 0 \\0 \\ 0 \\ 1
   \end{array}\right)
   =
   C
   \left(\begin{array}{c}
         B (\frac{1 - \cos \theta_y}{2})^2		e^{- 2 i \theta_z}
      \\ -B \frac{\sin\theta_y}{2} (1 - \cos \theta_y)	e^{-i \theta_z}
      \\ B (\sin \theta_y)^2 \sqrt{\frac{3}{8}} 
      \\ -B \frac{\sin\theta_y}{2} (1 + \cos \theta_y)	e^{i \theta_z}
      \\ B (\frac{1 + \cos \theta_y}{2})^2		e^{2 i \theta_z}
   \end{array}\right).
\label{eq:f2vec}
\end{equation}

\nidanHajime



\nidanOwari
\end{document}